\documentclass[twocolumn,tighten,twocolappendix]{aastex6}

\usepackage{amsmath}
\usepackage{graphicx}
\usepackage{epstopdf}
\usepackage{amssymb}
\usepackage{extarrows}
\usepackage{color}
\usepackage{bm}

\def\be{\begin{equation}}
\def\ee{\end{equation}}
\def\ba{\begin{eqnarray}}
\def\ea{\end{eqnarray}}
\def\go{\mathrel{\raise.3ex\hbox{$>$}\mkern-14mu
             \lower0.6ex\hbox{$\sim$}}}
\def\lo{\mathrel{\raise.3ex\hbox{$<$}\mkern-14mu
             \lower0.6ex\hbox{$\sim$}}}

\newcommand{\tildea}{{\tilde a}}
\newcommand{\bl}{\bm {\hat l}}
\newcommand{\bj}{\bm {\hat j}}
\newcommand{\der}{\text{d}}
\newcommand{\bL}{\bm L}
\newcommand{\bJ}{\bm J}
\newcommand{\bS}{\bm {\hat S}}
\newcommand{\hatS}{\hat S}
\newcommand{\ps}{\rm ps}

\def\red#1 {\textcolor{red}{#1}\ }   
\def\blue#1 {\textcolor{blue}{#1}\ }   

\begin{document}

\title{Solar Obliquity Induced by Planet Nine: Simple Calculation}

\author{Dong Lai}
\affil{Cornell Center for Astrophysics and Planetary Science, Department of Astronomy,
Cornell University, Ithaca, NY 14853}

\begin{abstract}
Bailey et al.~(2016) and Gomes et al.~(2016) recently suggested that
the 6 degree misalignment between the Sun's rotational equator and the
orbital plane of the major planets may be produced by the forcing from the
hypothetical Planet Nine on an inclined orbit. Here we present a
simple but accurate calculation of the effect, which provides a clear
description of how the Sun's spin orientation depends on the property
of Planet Nine in this scenario.
\end{abstract}


\section{Introduction}

Batygin \& Brown (2016) showed that a hypothetical planet (``Planet
Nine'') in the outer solar system can explain several otherwise
intriguing orbital properties of distant KBOs. Recently, Bailey et
al.~(2016) and Gomes et al.~(2016) suggested that Planet Nine, which
has an inclined orbit relative to the orbital plane of the major
planets, may also be responsible for generating the $6^\circ$ solar
obliquity (the misalignment angle between the Sun's rotational equator
and the solar system invariant plane). These studies were based on
somewhat formal treatments and involved numerical integrations.  In
this note we present a pedestrian, but accurate calculation of the
solar obliquity generated by Planet Nine.  This calculation yields a
simple and transparent description of how the solar spin orientation
depends on the property of Planet Nine.

\section{Explicit Analytic Calculation}

Batygin \& Brown (2016) showed that to explain the spatial clustering
of the orbits of distant KBOs, Planet Nine (labeled ``$p$'') must have
mass $m_p=(5-20)m_\oplus$, perihelion distance $q_p=a_p(1-e_p)\sim
250$~au, significant eccentricity ($e_p\go 0.5$) and tens of degrees
of orbital inclination with respect to the solar system invariant plane.  The
angular momentum of Planet Nine is $\bL_p = L_p \bl_p$ (where $\bl_p$ is 
the unit vector), with
\be
L_p
=0.276\,L_J\left({m_p\over 10\,m_\oplus}\right)
\left({\tildea_p\over 400\,{\rm au}}\right)^{1/2}(1-e_p^2)^{1/4},
\label{eq:Lp}\ee
where $L_J$ is the orbital angular momentum of Jupiter and 
we have defined the ``effective'' semi-major axis $\tildea_p\equiv a_p\sqrt{1-e_p^2}$.

Planet Nine exerts a torque on each of the ``canonical'' planets (labeled ``$j$'',
from Mercury to Neptune); this torque tends to induce a retrograde 
nodal precession of $\bl_j$ (the orbital angular momentum unit vector of planet $j$)
around $\bl_p$ at the characteristic rate
\be
\Omega_{jp}={3m_p\over 4M_\star}\left({a_j\over \tildea_p}\right)^3 n_j,
\ee
where $a_j,\,n_j$ are the semi-major axis and mean motion of planet j.
Note that $\Omega_{jp}$ depends on $a_j$, so each planet has 
a different $\Omega_{jp}$.
However, since the precession frequency due to mutual planet-planet interactions is
much larger than the differential $\Omega_{jp}$, all the canonical planets are
strongly coupled, with their angular momentum axes aligned to each other, i.e.,  
$\bl_j=\bl$ [see Lai \& Pu (2016) for a precise calculation of 
the mutual inclinations induced by an inclined external perturber].
The orbital angular momentum unit vector $\bl$ of the canonical solar
system planets then evolves according to the equation
\be
{\der\bl\over \der t}=\Omega_L \cos\theta_p \,(\bl\times\bl_p)
={J\over L_p}\Omega_L\cos\theta_p \,(\bl\times\bj),
\label{eq:dl}\ee
where $\theta_p$ is the inclination of Planet Nine
($\cos\theta_p=\bl\cdot\bl_p$), $\bJ$ is the total angular momentum
\be
\bJ=J\,\bj=\bL +\bL_p=L\,\bl +L_p\,\bl_p,
\ee
with $L=\sum_j L_j= 1.624 L_J$ (note that the spin angular momentum of the Sun,
$S_\star\sim 0.01 L_J$, is much smaller), and 
\ba
&& \Omega_L={\sum_j L_j\Omega_{jp}\over L} =2.74\Omega_{Jp}\nonumber\\
&&\quad ~={2\pi\over 87.5\,{\rm Gyr}}\left({m_p\over 10m_\oplus}\right)
\left({\tildea_p\over 400\,{\rm au}}\right)^{-3}.
\ea

The spin axis $\bS_\star$ (unit vector) of the Sun evolves due to the 
torques from all planets,
\be
{\der\bS_\star\over \der t}=\Omega_{\star\ps}
\cos\theta_{\rm sl} (\bS_\star\times\bl),
\label{eq:dS}
\ee
where $\theta_{\rm sl}$ is the angle between $\bS_\star$ and $\bl$ and 
the characteristic spin precession frequency is given by
\be
\Omega_{\star\ps}=\sum_j \Omega_{\star j}=
\sum_j{3k_{q\star}\over 2k_\star}\left( {m_j\over M_\star}\right)
\left({R_\star\over a_j}\right)^3\Omega_\star.
\ee
Here $\Omega_\star=2\pi/P_\star$ is the angular frequency of the Sun,
and $k_\star$,$k_{q\star}$ are defined through the Sun's moment of inertia and
quadrupole moment: $I_3=k_\star M_\star R_\star^2$ and $I_3-I_1
=k_{q\star}{\hat\Omega}_\star^2 M_\star R_\star^2$, with $\hat\Omega_\star=
\Omega_\star (GM_\star/R_\star^3)^{-1/2}$. Normalizing to the values 
$k_\star\simeq 0.06$ and $k_{q\star}\simeq 0.01$ (corresponding to 
$J_2=k_{q\star}\hat\Omega_\star^2\simeq 2.2\times 10^{-7}$; Mecheri et al.~2004), we 
find
\be
\Omega_{\star\ps}=2.88\, \Omega_{\star J}
={2\pi\over 55.8\,{\rm Gyr}}\, \lambda_\star \left({P_\star\over 10\,{\rm day}}\right)^{-1},
\ee
where $\lambda_\star\equiv 6k_{q\star}/k_\star\simeq 1$.

Equations (\ref{eq:dl}) and (\ref{eq:dS}) completely determine the evolution of
the spin axis of the Sun. In Eq.~(\ref{eq:dl}) we have neglected
the torque from the Solar spin on $\bl$, and in Eq.~(\ref{eq:dS}) we have neglected
the torque from $m_p$ on $\bS_\star$; both are excellent approximations.

To solve $\bS_\star(t)$ analytically, we note that 
Eqs.~(\ref{eq:dl}) implies that $\bl$ precesses around the constant unit vector $\bj$
at the rate $-(J/L_p)\Omega_L\cos\theta_p$. We transform 
Eq.~(\ref{eq:dS}) into the frame corotating with $\bl$, giving
\be
\left({\der\bS_\star\over \der t}\right)_{\rm rot}
=\left({J\over L_p}\,\Omega_L\cos\theta_p\,\bj -
\Omega_{\star\ps}\cos\theta_{\rm sl}\,\bl \right) \times \bS_\star.
\label{eq:dSrot}\ee
In this rotating frame, $\bj$ and $\bl$ are constant in time, 
and for $\theta_{\rm sl}\ll 1$ 
and constant $P_\star$,\footnote{The rotation rate of the Sun decreases 
over time due to magnetic braking. The structure parameter $\lambda_\star$ also
changes due to stellar evolution. We ignore these complications
and treat $\lambda_\star/P_\star$ as a free parameter.}
Eq.~(\ref{eq:dSrot}) describes a uniform rotation of 
$\bS_\star$ around a fixed axis (see Lai 2014). 
We set up a Cartesian coordinate system where $\bl={\hat z}$ and 
$\bl_p=-(\sin\theta_p) {\hat y}+ (\cos\theta_p) {\hat z}$ (so that the polar
and azimuthal angles of $\bl_p$ is $\theta_p$ and $\phi_p=270^\circ$).
In this coordinate system, Eq.~(\ref{eq:dSrot}) reduces to
\ba
&& {\der\hatS_{\star x}\over \der t}=-\Omega_y \hatS_{\star z} +\Omega_z 
\hatS_{\star y},\label{eq:Sxy1}\\
&&{\der\hatS_{\star y}\over \der t}=-\Omega_z \hatS_{\star x},
\label{eq:Sxy2}\ea
where 
\ba
&&\Omega_y = \Omega_L\sin\theta_p\cos\theta_p,\\
&&\Omega_z \simeq \Omega_{\star\ps} -\Omega_L\cos\theta_p \left({L\over L_p}+\cos\theta_p
\right).
\ea
For $\hatS_{\star z}\simeq 1$ (consistent with $\theta_{\rm sl}\ll 1$), 
Eqs.~(\ref{eq:Sxy1})-(\ref{eq:Sxy2}) 
can be solved (assuming that $\bS_\star$ is aligned with $\bl$ 
at $t=0$):
\be
\hatS_{\star x}=-{\Omega_y\over\Omega_z}\sin\Omega_z t,
\quad
\hatS_{\star y}={\Omega_y\over\Omega_z}(1-\cos\Omega_z t).
\ee
Thus the polar and azimuthal angles of $\bS_\star$ are given by
\ba
&& \theta_{\rm sl}\simeq (\hatS_{\star x}^2+\hatS_{\star y}^2)^{1/2}=
\left|{2\Omega_y\over\Omega_z}\sin{\Omega_z t\over 2}\right|,\label{eq:tht}\\
&& \phi_\star=\tan^{-1}{\hatS_{\star y}\over\hatS_{\star x}}=
\pi-{\Omega_z t\over 2}.
\label{eq:phistar}
\ea

\section{Dependence and Constraint on Planet Nine Parameters}

\begin{figure}
\vskip -1.4cm
\centering
\includegraphics[scale=0.43]{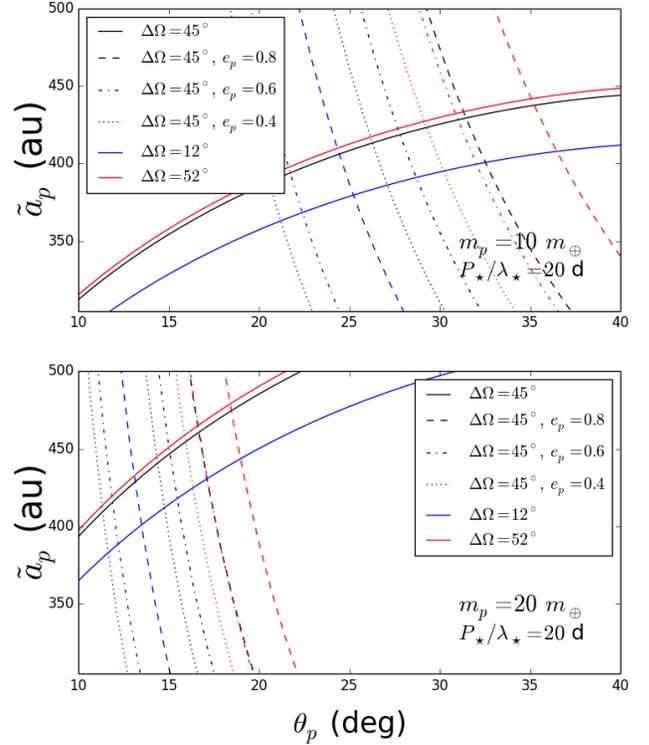}
\vskip -0.4cm
\caption{Parameters of Planet Nine required to produce the solar
  obliquity $\theta_{\rm sl}=6^\circ$ and the relative longitude of
  ascending node $\Delta\Omega$ (between Planet Nine and the solar
  equator). The effective semi-major axis of Planet Nine ${\tilde
    a}_p\equiv a_p (1-e_p^2)^{1/2}$ (where $a_p$ is the semi-major
  axis) is shown as a function of $\theta_p$, the inclination of
  Planet Nine relative to the orbital plane of the canonical solar
  system planets.  The mass of Planet Nine is set to $m_p=10m_\oplus$
  (upper panel) and $20m_\oplus$ (lower panel) and the solar rotation
  parameter is set to $P_\star/\lambda_\star=20$~days. The three solid
  lines depict Eq.~(\ref{eq:apcon}) with three values of
  $\Delta\Omega$ (covering the allowed range, $12^\circ-52^\circ$).
  The dashed and dotted lines depict Eq.~(\ref{eq:Lpcon}) for
  different $\Delta\Omega$ and $e_p$ (the eccentricity of Planet
  Nine). The intersect of a solid line and a corresponding
  dashed/dotted line of the same color marks the values of ${\tilde
    a}_p$ and $\theta_p$ (with the corresponding $m_p$ and $e_p$)
  required to generate the observed $\theta_{\rm sl}=6^\circ$ and
  $\Delta\Omega$.  }
\label{fig1}
\end{figure}

\begin{figure}
\vskip -1.4cm
\centering
\includegraphics[scale=0.43]{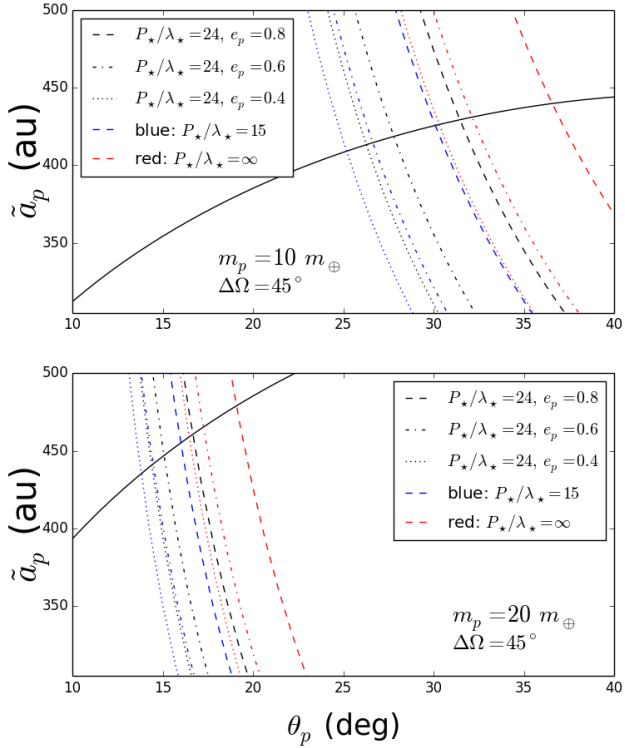}
\vskip -0.4cm
\caption{Similar to Fig.~1, except that the different curves correspond to
  different values of the solar rotation parameter $P_\star/\lambda_\star=24$~days
  (black; the current solar rotation period), 15~days (blue) and
  $\infty$ (red; implying that the spin axis of the Sun is constant in
  time, the limit considered by Gomes et al.~2016), all for
  $\Delta\Omega=45^\circ$. Note that the solid line (depicting
  Eq.~\ref{eq:apcon}) does not depend on $P_\star/\lambda_\star$.  }
\label{fig2}
\end{figure}

\begin{figure}
\vskip -0.3cm
\centering
\includegraphics[scale=0.43]{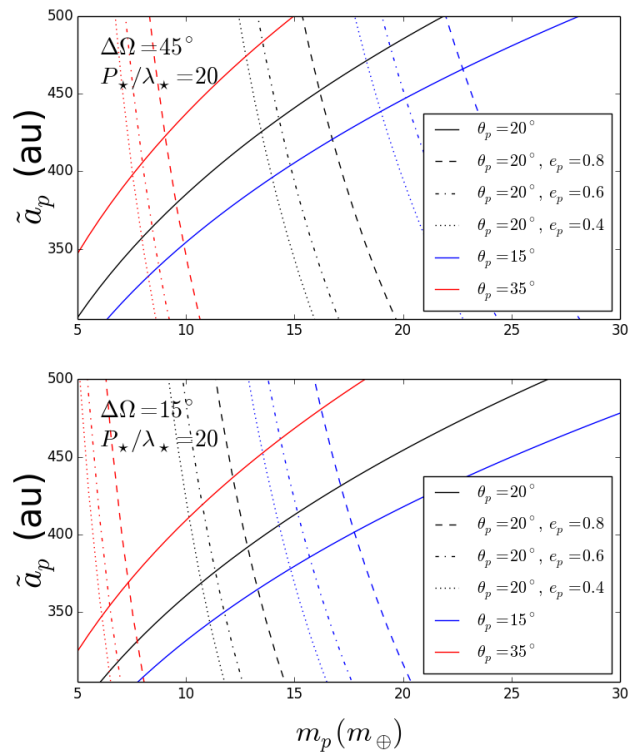}
\vskip -0.4cm
\caption{Similar to Fig.~1, except that ${\tilde a}_p$ is shown as a function 
of $m_p$ for several values of $\Delta\Omega$, $\theta_p$ and $e_p$,
as indicated.
}
\label{fig3}
\end{figure}

For a given $P_\star/\lambda_\star$, the values of $\theta_{\rm sl}$ and
$\phi_\star$ at $t=4.5$~Gyrs depend on the parameters of Planet Nine 
(${\tilde a}_p, e_p, m_p, \theta_p$) through the combination
of two frequencies, $\Omega_y$ and $\Omega_z$.
Batygin \& Brown (2016) suggested that the longitude of ascending node of Planet Nine
(relative that of the solar equator),
$\Delta\Omega\equiv \phi_p-\phi_\star$, is about $45^\circ$ and
ranges from $12^\circ$ to $52^\circ$. 
To produce this $\Delta\Omega$ 
and $\theta_{\rm sl}=6^\circ$ over time 4.5~Gyrs, the parameters of Planet Nine
must satisfy the following conditions, derived from 
Eqs.~(\ref{eq:tht})-(\ref{eq:phistar}):
\ba
&&{\tilde a}_p\simeq 462\,\left({m_p\over 10\,m_\oplus}\,{\sin 2\theta_p\over
{\hat\theta}_{\rm sl} f}\right)^{1/3}{\rm au},\label{eq:apcon}\\
&&{L\over L_p}\simeq \left[ 15 g+4.84 \lambda_\star\!\left(\!
{P_\star\over 10\,{\rm d}}\!\right)^{\!\!-1}\right]{\sin\theta_p\over {\hat\theta}_{\rm sl} f}-\cos\theta_p,\label{eq:Lpcon}
\ea
where $L=1.624L_J$ and $L_p$ is given by Eq.~(\ref{eq:Lp}), and 
we have defined
\be
{\hat\theta}_{\rm sl}\equiv {\theta_{\rm sl}\over 6^\circ},\quad
g\equiv {\pi/2 -\Delta\Omega\over \pi/4},\quad
f\equiv \left|{\pi/2-\Delta\Omega\over\cos\Delta\Omega}\right|.
\ee

Figures 1-3 illustrate the parameter space of Planet Nine required 
produce $\theta_{\rm sl}=6^\circ$ and $12^\circ <\Delta\Omega <52^\circ$.
Figure 1 shows the effective semi-major axis ${\tilde a}_p$ as a function of
$\theta_p$ for several values of planet mass and eccentricity,
assuming an ``averaged'' $P_\star/\lambda_\star$ or 20~days. Figure 2 illustrates
how the result depends on the solar rotation parameter $P_\star/\lambda_\star$.
Figure 3 shows ${\tilde a}_p$ as a function of $m_p$ for several values of $\theta_p$ 
and $e_p$. In general, a larger $m_p$ requires a smaller $\theta_p$, with a modest
change in ${\tilde a}_p$. There exists a minimum value of $\theta_p$, as indicated
by Eq.~(\ref{eq:Lpcon}). In all cases, ${\tilde a}_p$ lies in the range between
340~au and 480~au in order to produce the desired solar spin orientation.

\section*{Acknowledgments}
This work has been supported in part by
NASA grants NNX14AG94G and NNX14AP31G, and
a Simons Fellowship from the Simons Foundation.


\end{document}